\documentclass[pra,aps,showpacs, twocolumn]{revtex4-1}
\usepackage{mathtext}
\usepackage[T2A]{fontenc}
\usepackage[cp1251]{inputenc}
\usepackage{epsf}
\usepackage{psfrag}
\usepackage{graphicx}
\usepackage{amssymb,amsmath}
\usepackage{xcolor}

\begin{document}

\title{Single atom optical gate and single photon source based on magnetooptical effects in a waveguide}
\author{A. S. Kuraptsev}
\email[]{aleksej-kurapcev@yandex.ru}
\affiliation{\small Peter the Great St. Petersburg Polytechnic University, 195251, St. Petersburg, Russia}
\affiliation{\small Institute for Analytical Instrumentation of RAS, 198095, St. Petersburg, Russia}
\author{I. M. Sokolov}
\email[]{sokolov_im@spbstu.ru}
\affiliation{\small Peter the Great St. Petersburg Polytechnic University, 195251, St. Petersburg, Russia}

\date{\today}

\sloppy



\begin{abstract}
We have discovered abnormally strong influence of the magnetic field on the optical properties of atomic ensemble confined in a waveguide. We demonstrate qualitative changes in the character of spontaneous emission and single-atom susceptibility. Based on the revealed effects, we propose a new scheme of true single photon source. Furthermore, we propose the highly-efficient optical gate using just one atom.
\end{abstract}
\maketitle

\section{Introduction}
Rapid progress in quantum information science requires compact and effective sources of single photons, nanodetectors and optical gates on the single photon level. For this purpose, extensive efforts are directed on the design of quantum systems with desired optical properties and the control over these properties in a real time \cite{intro1,intro2}.

Waveguide QED provides an excellent platform to handle this task, in particular, in terms of quantum information protocols. On the basis of atomic systems coupled with a waveguide, single-photon switches \cite{waveduide_group3_3,waveduide_group3_4,waveduide_group3_5}, routers \cite{waveduide_group3_6}, transistors \cite{waveduide_group3_7,waveduide_group3_8,waveduide_group3_9}, frequency comb generators \cite{waveduide_group3_10}, and single-photon frequency converters \cite{waveduide_group3_11} have been proposed.

When coupled with a waveguide, the entanglement between different atoms in an ensemble significantly increases owing to the guided modes. This understanding gave rise to a number of experiments with laser cooled atoms prepared in a nanofiber trap \cite{nanofiber1,nanofiber2,nanofiber3}. Afterwards, it allowed one to realize an efficient Bragg mirror using $\sim10^{3}$ atoms achieving a reflectance of up to $75\%$ \cite{nanofiberexp1,nanofiberexp2}.

Here we show that it is possible to realize an efficient and controllable Bragg mirror, which allows one to change the reflectance form 0 to 1 using just one atom if it is placed \emph{inside} the waveguide.

Generally, unique properties of atoms in a cavity or waveguide are caused by the fact that the spatial structure of the modes of the electromagnetic field in confined geometries is modified in comparison with free space. Consequently, the coupling between the atoms and the electromagnetic field changes, which, in turn, changes the character of any kind of the electromagnetic interaction. It turns out that in the case when the atom is located inside the waveguide, the coupling constant significantly exceed one in the case when the same atom is located in the outer space of a waveguide close to its surface.

Moreover, the modification of spatial characteristics of the vacuum modes of the electromagnetic field leads to the changes in the fundamental properties of atoms, in particular, the spontaneous emission. In 1974, V. P. Bykov published the pioneering work, which has demonstrated strong suppression of the spontaneous emission of an excited atom in a one-dimensional periodic structure with a band spectrum due to the appearance of the long-lived dark state \cite{Bykov}. Later, D. Kleppner reported the suppression of the spontaneous emission of an excited atom in a waveguide under the conditions when the transition frequency is less than the cutoff frequency of a waveguide \cite{Kleppner}. Actually, similar situation, when the non-decaying dark state of an atom is formed, takes place in photonic band gap crystals. It has been reported in a number of papers and summarized, for example, in the review \cite{Lambropoulos}.

In this paper, we report that non-decaying dark state of an excited atom in a waveguide can be controllable by external magnetic field. It has been found that strong magnetic field cancels the effect of incomplete spontaneous decay, which occurs for excited atom in a single-mode waveguide due to polarization selection \cite{PRA2020waveguide}. On this basis, we suggest the new scheme of single photon source. Using magnetic field we propose to prepare an atom in a specific excited state, which can be represented as a superposition of the decaying part and the non-decaying dark state. After the magnetic field is switched off, with time, the atom turns out to be in the long-lived dark state. In a desired moment of time, the magnetic field is switched on again making the atom to emit a photon.

\section{BASIC ASSUMPTIONS AND APPROACH}
The system considered here employs an ensemble of point-like motionless atoms in a waveguide under the action of external magnetic field, Fig. 1. This model is excellent for ensembles of impurity atoms embedded in a transparent dielectric under low temperatures that, therefore, provide a fantastic and practically realizable playground for testing the theory \cite{Naumov1,Naumov2}. In the optical domain, waveguide is usually associated with optical fiber doped with active impurities or hollow core gas filled fiber \cite{Bufetov1,Bufetov2}. Applying the proposed scheme to the microwave domain, waveguide is represented by the metal tube, and probe radiation should be resonant to the microwave transition between Rydberg states of the impurity atoms embedded in a solid transparent glass or crystal matrix \cite{Abmann}. Nowadays, waveguide platform also attracts a special attention in quantum optics with artificial atoms \cite{Astafiev}.

\begin{figure}\center
	\includegraphics[width=7cm]{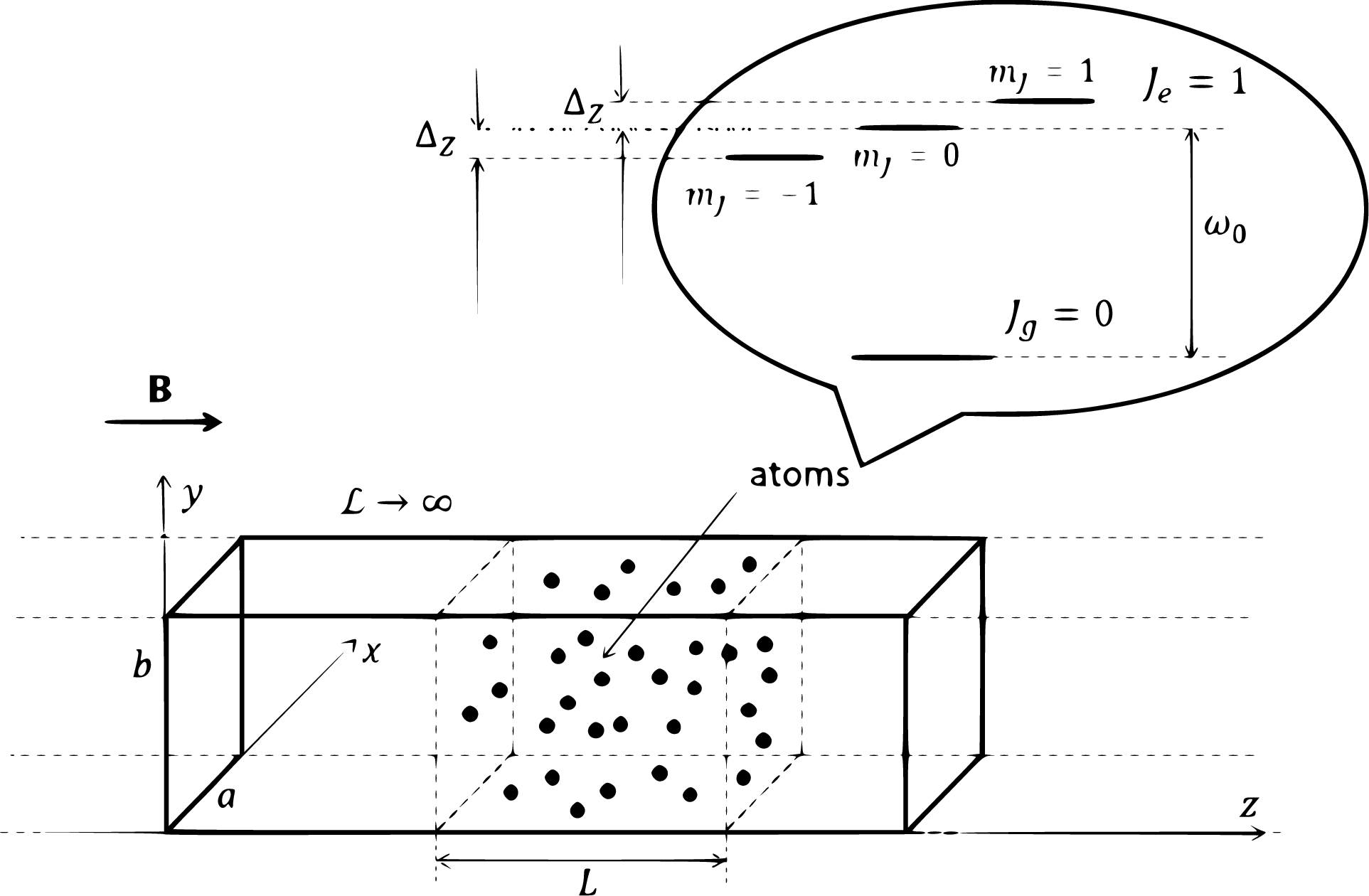}
	\caption{\label{fig:one}
			Sketch of the waveguide, the atomic ensemble inside it and the permanent magnetic field. The inset illustrates the model scheme of atomic energy levels in a magnetic field. $\omega_{0}$ is the resonant transition frequency of a free atom; $\Delta_{Z}$ is Zeeman splitting of the excited state triplet, which is proportional to the absolute value of the magnetic field strength.}\label{f1}
\end{figure}

We assume that the atoms are equal, having a nondegenerate ground state $|g_{i}\rangle$ with energy $E_{g}$ and the total angular momentum $J_{g}=0$ and an excited state $|e_{i}\rangle$ with $J_{e}=1$ and natural free space linewidth $\gamma_{0}$ (the index $i=1,...,N$
denotes quantities corresponding to the atom $i$ among $N$
atoms). The excited state is represented by three Zeeman sublevels $|e_{i,m_{J}}\rangle$, which differ by the angular momentum projection on the quantization axis $z$ -- $m_{J}=-1,0,1$. For convenience, let us choose $z$ axis coinciding with the axis of a waveguide. In the absence of external magnetic field, the excited state is triply degenerate having the energy $E_{e}=E_{g}+\hbar\omega_{0}$. Permanent magnetic field directed along $z$ axis, $\textbf{B}$, cancels the degeneracy of the exited state, so Zeeman splitting $\Delta_{Z}$ appears. The presence of constant magnetic field in the considered geometry satisfies Maxwell's equations for the inner space of a waveguide. For simplicity, let us restrict ourselves by the linear Zeeman effect. In this case, the state $m_{J}=-1$ has the energy $E_{e_{m_{J}=-1}}=E_{g}+\hbar(\omega_{0}-\Delta_{Z})$; $m_{J}=0$ -- $E_{e_{m_{J}=0}}=E_{g}+\hbar\omega_{0}$; $m_{J}=1$ -- $E_{e_{m_{J}=1}}=E_{g}+\hbar(\omega_{0}+\Delta_{Z})$, see the inset in Fig. 1.

Assuming the walls of a waveguide to be perfectly conductive (i.e. neglecting the absorption), we can write the non-steady-state Schrodinger equation for the wave function of the joint system, which consists of the atoms and the electromagnetic field in a waveguide, including vacuum reservoir. This system is described by the following Hamiltonian:
\begin{eqnarray}\label{equation1}
  \widehat{H}&=&\sum_{i=1}^{N}\sum_{m_{J}=-1}^{1}E_{e_{i,m_{J}}}|e_{i,m_{J}}\rangle\langle e_{i,m_{J}}| \nonumber \\
  &+&\sum_{\textbf{k},\alpha}\hbar\omega_{k}\left(\widehat{a}_{\textbf{k},\alpha}^{\dagger}\widehat{a}_{\textbf{k},\alpha}+\frac{1}{2}\right)-\sum_{i=1}^{N}\widehat{\textbf{d}}_{i}
  \cdot\widehat{\textbf{E}}\left(\textbf{r}_{i}\right) \nonumber \\
  &+&\frac{1}{2\epsilon_{0}}\sum_{i\neq j}^{N}\widehat{\textbf{d}}_{i}\cdot\widehat{\textbf{d}}_{j}\delta\left(\textbf{r}_{i}-\textbf{r}_{j}\right)+g_{e}\mu_{B}\textbf{B}\cdot\textbf{J}_{e},
\end{eqnarray}
where the first two terms correspond to noninteracting
atoms and the electromagnetic field in an empty waveguide, respectively; the
third term describes the interaction between the atoms and
the electromagnetic field in the dipole approximation; the fourth, contact
term, ensures the correct description of the electromagnetic
field radiated by the atoms; and the last, fifth term, describes the action of the constant magnetic field on the atoms. In Eq. (\ref{equation1}), $\widehat{a}_{\textbf{k},\alpha}^{\dagger}$ and $\widehat{a}_{\textbf{k},\alpha}$ are the
operators of creation and annihilation of a photon in the corresponding mode, $\omega_{k}$ is the photon frequency, $\widehat{\textbf{d}}_{i}$ is the dipole operator of the atom $i$, $\widehat{\textbf{E}}\left(\textbf{r}\right)$ is the electric displacement vector in a waveguide, $\textbf{r}_{i}$ is the position of the atom $i$, $\epsilon_{0}$ is the vacuum permittivity, $\mu_{B}$ is the Bohr magneton, and $g_{e}$ is the Lande factor of the excited state.

To handle the problem mathematically, we apply quantum microscopic approach developed previously for dense and cold atomic gases \cite{SokolovJETP2011} and adopted afterwards for atomic ensembles confined in a waveguide \cite{PRA2020waveguide,PRA2022waveguide,PRA2023waveguide}. In the framework of this approach, we decompose the wave function of the combined atomic-field system in a set of eigenfuctions of non-interacting atoms and free electromagnetic field. Using this technique, we convert the Schrodinger equation with Hamiltonian (\ref{equation1}) to the infinite set of equations for the quantum amplitudes of the combined atomic-field system. Then, restricting ourselves by the regime of linear optics, we are able to express the quantum amplitudes of field subsystem via the amplitudes of the one-fold atomic excited states, $b_{e_{i,m_{J}}}$. Thus, we obtain a finite set of linear equations for the amplitudes of the one-fold atomic excited states.

General formalism allows us to consider both time-dependent problems and stationary ones. In time-dependent tasks, the problem statement always assumes initial conditions. Accounting for the task considered in this paper, we will focus on the case when one atom is excited at the initial time, other atoms are in their ground states and the electromagnetic state is in the vacuum state.
In this case, one obtains a system of equations for the amplitudes $b_e$ of one-fold atomic excited states with the coupling
between atoms caused by the dipole-dipole interaction. For Fourier components $b_{e}(\omega)$ we have
(at greater length see Ref. \cite{SokolovJETP2011})
\begin{equation}\label{set1}
\sum_{e'}\bigl[(\omega-\omega_{e})\delta_{ee'}-\Sigma_{ee'}(\omega)\bigl]b_{e'}(\omega)=i\delta_{e s}.
\end{equation}
Hereafter, the index $s$ as well as the indexes $e$ and $e'$ contain information both about the number of atom and about specific atomic sublevel excited in the corresponding state (additional indexes $i,m_{J}$ will be omitted afterwards). The index $s$ indicates populated sublevel of the initially excited atom. The matrix $\Sigma_{ee'}(\omega)$ describes both spontaneous decay and photon exchange between the atoms. This is the key quantity in the microscopic theory. Concerning to an ensemble of atoms in a waveguide, detailed derivation of this matrix can be found in the Appendix in Ref. \cite{PRA2020waveguide} and, therefore, will not be reproduced here. By the inverse Fourier transform, one can get the dynamics of the quantum amplitudes in a time domain, $b_{e}(t)$.
Actually, the initial value problem can be considered in the general form, including the case of distributed atomic excitation.

In stationary tasks, the problem statement usually assumes external probe light illuminating the atomic ensemble. In this case we are able to analyze the properties of scattered radiation, calculate the transmittance and reflectance. The bridge from the time-domain problem to the stationary regime goes throw the consideration of the point source, which is implemented by the specific atom called "source atom". Eq. (\ref{set1}) accounts for arbitrary spatial location of initially excited atom. Passing to the stationary regime, we should consider initially excited source atom far from the atomic ensemble generating the probe wave. Source atom has the same level structure as atoms of the ensemble but different resonant transition frequency, $\omega_{s}$, and its natural linewidth $\gamma_{s}\rightarrow 0$. The latter simulates monochromatic probe light. Following the technique described in detail in Ref. \cite{PRA2022waveguide}, we consider $t\rightarrow\infty$ assuming $\gamma_{s}t\ll 1$.
Thus, after these limiting passages, we obtain the final expression for stationary quantum amplitudes:
\begin{equation}\label{set2}
b_{e}^{st}(t)=\exp(-i\omega_{s}t)\sum_{e'\neq s}R_{ee'}(\omega_{s})\Sigma_{e's}(\omega_{s}),
\end{equation}
where $R_{ee'}(\omega)$ is a resolvent operator of the considered multi-atomic ensemble, which is defined as
$R_{ee'}(\omega)=[(\omega-\omega_{e})\delta_{ee'}-\Sigma_{ee'}(\omega)]^{-1}$. Here, the index $s$ refers to the source atom. Note that expression (\ref{set2}), in fact, represents a set of linear algebraic equations.

\section{RESULTS AND DISCUSSION}
\subsection{Single photon source}
At the beginning, we consider spontaneous decay dynamics of a single atom confined in a waveguide under the action of external magnetic field.
Figure 2 shows the dynamics of the excited state probability for both Zeeman sublevels $m_{J}=-1$ and $m_{J}=1$ of the atom located in a single-mode waveguide with the transverse sizes $k_{0}a=4$ and $k_{0}b=2$; $k_{0}$ is the wavenumber of radiation resonant to the transition of a free atom, $k_{0}=\omega_{0}/c$. In such a waveguide, only $\text{TE}_{10}$ mode can propagate at long distances as an oscillating wave at the frequency $\omega_{0}$. Without the restriction of generality, we assume that the atom is located in the center of the cross section of a waveguide, having the coordinates $x_{a}=a/2$, $y_{a}=b/2$. At the initial time, only one Zeeman sublevel of the excited state of this atom, $m_{J}=-1$, is populated.
\begin{figure}\center
	\includegraphics[width=7cm]{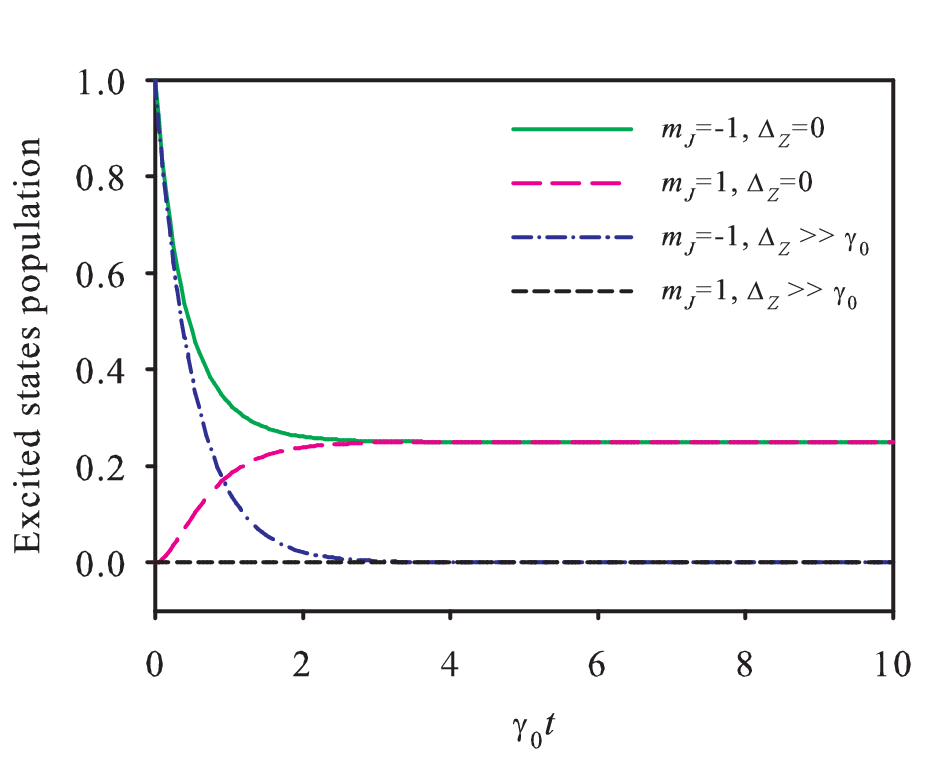}
	\caption{\label{fig:two}
			Time dependence of the excited states population. The sizes of the cross section of the waveguide are $k_{0}a=4$, $k_{0}b=2$; the position of the atom $x_{a}=a/2$, $y_{a}=b/2$. At the initial time, only one sublevel $m_{J}=-1$ of the excited state is populated.}\label{f2}
\end{figure}

The result shown in Fig. 2 is given both for the case of strong magnetic field resulting in the Zeeman splitting $\Delta_{Z}\gg \gamma_{0}$ and for the case of absence of the magnetic field. In the case when external magnetic field is absent, $\textbf{B}=0$ ($\Delta_{Z}=0$), atomic excitation dynamics exhibits the effect of incomplete spontaneous decay, which was described in detail in Ref.  \cite{PRA2020waveguide}. This effect, in its essence, is caused by the appearance of the dark state $|X\rangle=(|m_{J}=-1\rangle-|m_{J}=1\rangle)/\sqrt{2}$ corresponding to linear orientation of the atomic dipole momentum along $x$ axis. Since the main guided mode of a waveguide, $\text{TE}_{10}$, has no $x$ component of the electric field, the state $|X\rangle$ is non-decaying.

In contrast, in the case of strong magnetic field resulting in large Zeeman splitting, $\Delta_{Z}\gg \gamma_{0}$ we see normal spontaneous decay -- the population of the sublevel $m_{J}=-1$ exponentially decrease from its initial value 1 approaching horizontal asymptote 0, whereas the sublevel $m_{J}=1$ is almost not populated. Thus, strong magnetic field suppresses the effect of incomplete spontaneous decay. The reason of this suppression can be explained by the following. Since $|X\rangle$ state is a superposition of $|m_{J}=-1\rangle$ and $|m_{J}=1\rangle$ states, it can be formed when the excited state triplet is degenerate. In strong magnetic field, the sublevels $|m_{J}=-1\rangle$ and $|m_{J}=1\rangle$ are separated in terms of frequency, so superposition state does not form.

The revealed effect allows us to propose a new scheme of single photon source. At first, the atom confined in a single-mode waveguide is excited at Zeeman sublevel $|m_{J}=-1\rangle$ in strong magnetic field using resonant $y$-polarized probe guided mode $\text{TE}_{10}$. Then, magnetic field is switched off resulting the degeneracy of the excited state triplet followed by incomplete spontaneous decay of $|m_{J}=-1\rangle$ state. As the result of this process, with time, the atom turns out to be in the long-lived non-decaying state $|X\rangle$. On demand, the magnetic field is switched on again, which cancels the dark state $|X\rangle$ and, consequently, the atom emits a photon.

The advantage of the proposed scheme is that the emitted photon is prepared in the Fock state, unlike commonly used single photon sources based on weakened laser radiation, which, in fact, generate radiation in the coherent state with the average number of photons much less that unity. Herewith, the time moment of photon emission is controllable (within the accuracy determined by $1/\gamma_{0}$). Another feature of the proposed scheme is that it does not require any nonlinearities in contrast with previous proposals of Fock state generation \cite{Kilin1995,Dani2003,Palmer2017,Ma2023}.

The performance of the proposed single photon source requires fast switching on/ off of the magnetic field. Specifically, the time of transient process must be much less than the natural lifetime of the atomic excited states. The latter has typical order of magnitude $\sim 10 - 100$ ns in the case of allowed electro-dipole transitions. The most simple consideration of the supply circuit, which consists of the voltage source, the magnetic coil and the key switch gives us the relation between the time of transient process, $\tau_{tr}$, and saturated value of the magnetic field. On the basis of the Breit-Rabi formula we are able to evaluate the magnetic field sufficient to provide the Zeeman splitting $\Delta_{Z}>\gamma_{0}$, $B\sim 1$ mT. With realistic values of the source voltage and the parameters of the magnetic coil, one can obtain $\tau_{tr}\sim 1$ ns. It proves the realness of experimental implementation of the proposed scheme.

\subsection{Optical gate}
Associated effect that we have found out concerns the transmittance and reflectance of atomic ensemble or just one atom confined in a waveguide under the action of external magnetic field. Calculation of the transmission coefficient belongs to the class of stationary problems. Mathematical technique of this calculation was described in detailed in Ref. \cite{PRA2022waveguide} for the case when external magnetic field is absent. According to the problem statement, the atoms are illuminated by the monochromatic probe light in a single-mode waveguide. Probe radiation represents $\text{TE}_{10}$ mode, and its frequency is close to the atomic transition. The detector is located behind the atomic medium of length $L$ and measures the integral signal transmitted through the sample.

Here we analyze the dependence of the transmission coefficient on the magnetic field. Figure 3(a) demonstrates this dependence for the atomic ensemble with strong interatomic dipole-dipole interactions. Spatial distribution of atoms is random but uniform on average. This type of distribution is typical for the experiments with impurity atoms embedded in a solid transparent dielectric. The reflectance, $R$, is connected with the transmittance, $T$, by the energy conservation law -- $R+T=1$. The detuning of the probe frequency, $\delta$, is counted from the exact resonance of $J=0\leftrightarrow J=1, m_{J}=-1$ transition.

\begin{figure}\center
	\includegraphics[width=7cm]{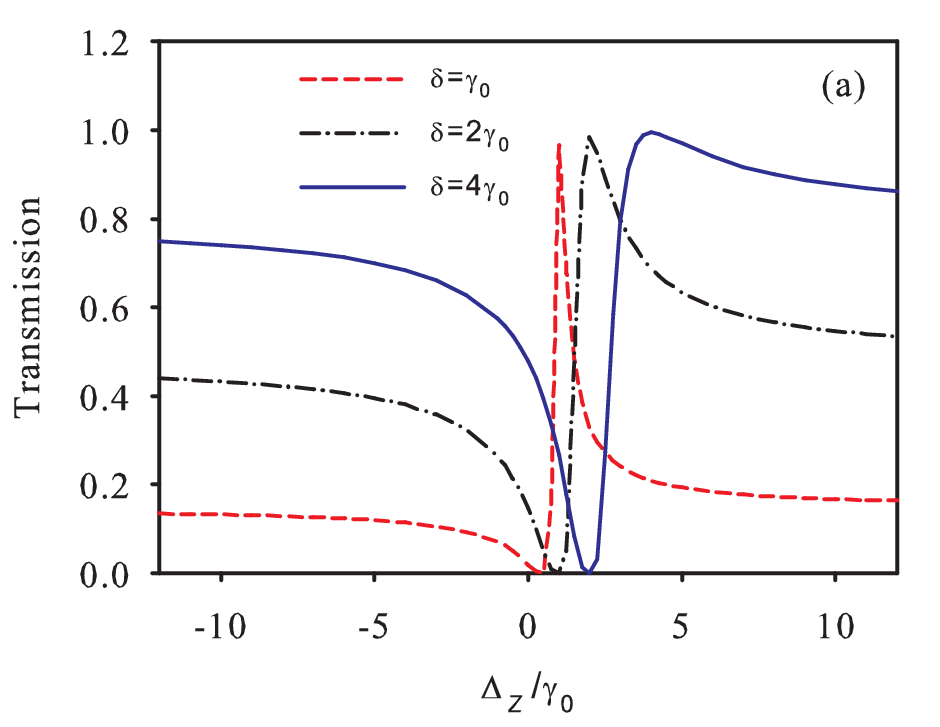}
	\includegraphics[width=7cm]{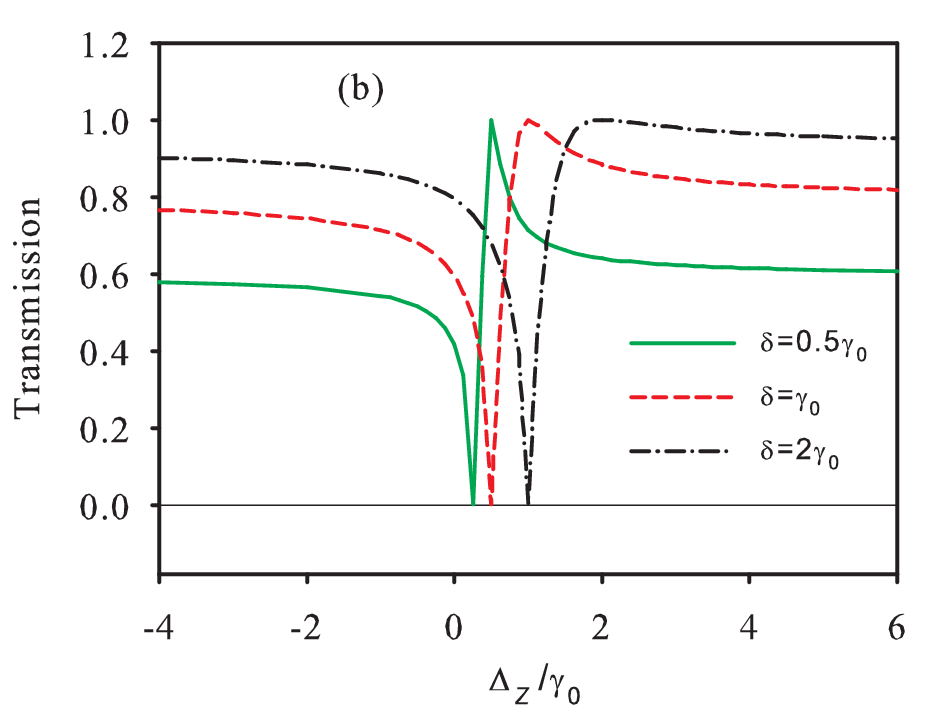}\\
	\caption{\label{fig:three}
			Transmission depending on the Zeeman splitting of the excited state triplet (proportional to the magnetic field). The sizes of the cross section of a waveguide $k_{0}a=4$, $k_{0}b=2$. The detuning of the probe frequency, $\delta$, is counted from the exact resonance of $J=0\leftrightarrow J=1, m_{J}=-1$ transition. Random (on average) atomic ensemble with the density of atoms $n\lambdabar^{3}=2\times 10^{-3}$, where $\lambdabar=1/k_{0}$, and longitudinal size $k_{0}L=750$ (a); single atom (the result is averaged over random atomic positions) (b).}\label{f3}
\end{figure}

Figure 3(a) shows that the adjustment of the magnetic field allows us to change the transmission coefficient in the diapason from 0 to 1. The calculation was performed for sufficiently large length of the sample -- $k_{0}L=750$, it is several times greater than the mean free path of a photon inside an atomic ensemble of the given density in free space. In the Figure 3(a) we see that the transmission is suppressed when Zeeman splitting is two times smaller than the detuning of the probe light whereas the transmission coefficient equals 1 when probe detuning coincide with Zeeman splitting. It means that when the probe frequency is tuned on the transition $J=0\leftrightarrow J=1, m_{J}=0$, atomic ensemble transmits all the probe power and reflects nothing. In contrast, when the probe frequency is tuned on the transition $J=0\leftrightarrow J=1, m_{J}=1$, the transmission coefficient equals 0, hence, the medium reflects all the probe radiation. This conclusion is quantitatively valid for the scheme of atomic levels considered here (as shown in the inset of Fig. 1). However, the feasibility to manage the transmittance and reflectance of the atomic ensemble in a wide diapason by the adjustment of the external magnetic field is naturally expected to be qualitatively valid for other schemes of levels.

Figure 3(b) shows the same dependence but calculated for the single atom in a waveguide. It demonstrates that the suppression of the transmittance, in its essence, can be interpretated as the single-atom effect. The character of the curves corresponding to the single atom in a waveguide qualitatively coincides with that corresponding to an ensemble. It justifies the feasibility to create an optical gate using just one atom.

The nature of the observed effects concerns the interference between two scattering channels, which are represented by Zeeman sublevels $m_{J}=-1$ and $m_{J}=1$ of the atom, as well as the interference between probe light and scattered secondary radiation. The suppression of the reflection, which is observed in the case when the probe frequency is tuned on the transition $J=0\leftrightarrow J=1, m_{J}=0$, is caused by the fact that the complex amplitudes of light scattered backward by the sublevels $m_{J}=-1$ and $m_{J}=1$ have identical absolute value and the phase difference $\pi$. Thus, in this case the destructive interference occurs. Zero transmission, which is observed in the case when the probe frequency is tuned on the transition $J=0\leftrightarrow J=1, m_{J}=1$, is caused by the destructive interference between probe wave and forwardly scattered secondary radiation by the resonant Zeeman sublevel $m_{J}=1$.

The case of a single atom located in a single-mode waveguide allows us to get an analytical solution of the set of equations (\ref{set2}). Our analysis shows that in the stationary regime, the quantum amplitude of the excited state $|m_{J}=-1\rangle$ reads
\begin{eqnarray}\label{equation2}
  b_{-}^{st}=A_{1}\frac{\delta-2\Delta_{Z}}{\bigl(\delta+i\frac{\gamma'}{2}\bigl)\bigl(\delta-2\Delta_{Z}+i\frac{\gamma'}{2}\bigl)+\bigl(\frac{\gamma'}{2}\bigl)^{2}}
\end{eqnarray}
and the quantum amplitude of the excited state $|m_{J}=1\rangle$ reads
\begin{eqnarray}\label{equation3}
  b_{+}^{st}=A_{1}\frac{\delta}{\bigl(\delta+i\frac{\gamma'}{2}\bigl)\bigl(\delta-2\Delta_{Z}+i\frac{\gamma'}{2}\bigl)+\bigl(\frac{\gamma'}{2}\bigl)^{2}},
\end{eqnarray}
where
\begin{eqnarray}\label{equation4}
  \gamma'=\gamma_{0}\frac{3\pi}{k_{0}^{2} a b \sqrt{1-\bigl(\frac{\pi}{k_{0}a}\bigl)^{2}}}\sin^{2}\Bigl(\frac{\pi}{a}x_{a}\Bigl).
\end{eqnarray}
The coefficient $A_{1}$ has the physical sense of the direct influence of the primary probe radiation on the sublevel $m_{J}=-1$ or $m_{J}=1$ separately and can be expressed as follows:
\begin{eqnarray}\label{equation5}
  A_{1}&=&-\frac{\pi i C}{k_{0}^{2} a b \sqrt{1-\bigl(\frac{\pi}{k_{0}a}\bigl)^{2}}}\sin\Bigl(\frac{\pi}{a}x_{a}\Bigl)\sin\Bigl(\frac{\pi}{a}x_{s}\Bigl) \nonumber \\
  &\times&\frac{3}{2}\gamma_{0}\exp\biggl(i|z_{a}-z_{s}|\sqrt{1-\Bigl(\frac{\pi}{k_{0}a}\Bigl)^{2}}\biggl),
\end{eqnarray}
where $x_{a}$ and $z_{a}$ are the coordinates of the atom; $x_{s}$ and $z_{s}$ are the coordinates of the point source, which emits probe radiation; $C$ is the normalization constant proportional to the amplitude of the probe wave. Note that Eq. (\ref{equation4}) accounts for the dependence of the spontaneous decay rate on the atomic spatial position.

Since we discuss the case of a single-mode waveguide, we can note that the electric field in $\text{TE}_{10}$ mode has only $y-$ nonzero component. Consequently, the electric component of the transmitted light power, $P_{t}$ is proportional to $|E_{t}^{y}|^{2}$. The transmitted signal is represented by the interference of the primary radiation of the source and secondary radiation scattered by the atom,
\begin{eqnarray}\label{equation6}
  E_{t}^{y}=\sqrt{\frac{\hbar k_{0}^{3}}{\gamma_{0}}}\Bigl(A_{2}-A_{3}(b_{-}^{st}+b_{+}^{st})\Bigl),
\end{eqnarray}
where the first term, $A_{2}$, is determined by the amplitude of the primary probe radiation at the observation point, and the second term describes secondary radiation scattered by the atom.
The coefficient $A_{2}$ from Eq. (\ref{equation6}) looks similar to $A_{1}$ if replace the coordinates of the atom, $x_{a}$ and $z_{a}$, by the coordinates of the observation point, $x_{d}$ and $z_{d}$,
\begin{eqnarray}\label{equation7}
  A_{2}&=&-\frac{\pi i C}{k_{0}^{2} a b \sqrt{1-\bigl(\frac{\pi}{k_{0}a}\bigl)^{2}}}\sin\Bigl(\frac{\pi}{a}x_{d}\Bigl)\sin\Bigl(\frac{\pi}{a}x_{s}\Bigl) \nonumber \\
  &\times&\frac{3}{2}\gamma_{0}\exp\biggl(i|z_{d}-z_{s}|\sqrt{1-\Bigl(\frac{\pi}{k_{0}a}\Bigl)^{2}}\biggl).
\end{eqnarray}
Finally, the coefficient $A_{3}$ is responsible for the impact of secondary waves scattered by the atom to the transmitted signal and can be expressed as follows:
\begin{eqnarray}\label{equation8}
  A_{3}&=&\frac{\pi i}{k_{0}^{2} a b \sqrt{1-\bigl(\frac{\pi}{k_{0}a}\bigl)^{2}}}\sin\Bigl(\frac{\pi}{a}x_{a}\Bigl)\sin\Bigl(\frac{\pi}{a}x_{d}\Bigl) \nonumber \\
  &\times&\frac{3}{2}\gamma_{0}\exp\biggl(i|z_{a}-z_{d}|\sqrt{1-\Bigl(\frac{\pi}{k_{0}a}\Bigl)^{2}}\biggl).
\end{eqnarray}

To calculate the reflected signal, $\textbf{E}_{r}$, one can use Eqs. (\ref{equation2}) -- (\ref{equation8}) if consider the observation point in front of the atom (specifically, $z_{d}<z_{a}$) and eliminate the coefficient $A_{2}$ from Eq. (\ref{equation6}), which refers to the impact of the primary probe radiation. The latter is connected with standard definition of the reflectance according to which, the reflected signal accounts only secondary scattered radiation. After that, we obtain the transmission coefficient, $T=\bar{P_{t}}/(\bar{P_{t}}+\bar{P_{r}})$, where $P_{r}$ is the electric component of the reflected light power, which is proportional to $|E_{r}^{y}|^{2}$; the upper bar means averaging over the transverse coordinates of the observation point, $x_{d}$ and $y_{d}$, which accounts the fact that the photodetector measures the integral signal. Normalization constant, $C$, is obviously reduced in the equations for the transmittance and reflectance.

Analytical calculations based on Eqs. (\ref{equation2}) -- (\ref{equation8}) reproduce the numerical result given by Fig. 3(b). In particular, in the case when $\delta=\Delta_{Z}$ (the probe frequency is tuned on the transition $J=0\leftrightarrow J=1, m_{J}=0$), from Eqs. (\ref{equation2}) and (\ref{equation3}) one clearly see that $b_{-}^{st}=-b_{+}^{st}$, hence, the reflected field $E_{r}^{y}=-A_{3}(b_{-}^{st}+b_{+}^{st})\sqrt{\hbar k_{0}^{3}/\gamma_{0}}=0$. Therefore, in this case we obtain $R=0$ and $T=1$. In the other case, when $\delta=2\Delta_{Z}$ (the probe frequency is tuned on the transition $J=0\leftrightarrow J=1, m_{J}=1$), we get $b_{-}^{st}=0$ and $b_{+}^{st}=-2i A_{1}/\gamma'$. Substituting here the expressions (\ref{equation4}) and (\ref{equation5}), one can ealily prove that $b_{+}^{st}A_{3}
=A_{2}$. Thus, according to Eq. (\ref{equation6}), we get $E_{t}^{y}=0$ and, consequently, $T=0$.

\section{CONCLUSION}
In conclusion, we have studied the optical properties both of a single atom and of an atomic ensemble with strong dipole-dipole coupling in a waveguide under the action of external magnetic field. It has been found that strong magnetic field cancels the effect of incomplete spontaneous decay, which takes place in the case of an atom confined in the single-mode waveguide in the absence of magnetic field. The revealed effect allows us to suggest the new scheme of single photon source. The associated effect that has been found is that magnetic field allows one to manage the transmittance and reflectance of atomic medium in the range from 0 to 1. We have proved that the suppression of the transmittance is essentially single-atom effect. It justifies the feasibility to create a highly efficient and controllable optical gate using just one atom.

Finally, we note that the effects described in this paper are caused by the specific three-dimensional structure of the modes of the electromagnetic field in a waveguide and, consequently, theoretical description of the discussed features requires precise geometric specification of a waveguide. It is an important point that differs this paper from a number of other theoretical papers, which employ a quasi-one-dimensional treatment avoiding the detailed profile of the field modes (see, for example, \cite{Domokos} and associated references). This is explained by the fact that the effect of incomplete spontaneous decay described in \cite{PRA2020waveguide} and used here is caused by polarization selection and, therefore, it represents an essentially three-dimensional effect.

\section*{Acknowledgments}
A.S.K. appreciates financial support from the
Foundation for the Advancement of Theoretical Physics and Mathematics ”BASIS” (Agreement 24-1-3-12-1). The results of the work were obtained using computational resources of Peter the Great St. Petersburg Polytechnic University Supercomputer
Center (http://www.scc.spbstu.ru).

\end{document}